\documentclass[12pt]{article}
\begin{document}
\title{A Comment on the Paper, ``Cosmic Background Radiation"}
\author{B.G. Sidharth\\
G.P. Birla Observatory \& Astronomical Research Centre\\
B.M. Birla Science Centre, Adarsh Nagar, Hyderabad - 500 063
(India)}
\date{}
\maketitle

In the paper \cite{bgscsf} and its updated version \cite{valluri} it was shown that a collection of Photons with nearly the same energy or frequency exhibit a condensation type of behaviour. To see what this means, our starting point is the formula for the average occupation number for photons of momentum $\vec{k}$ for all polarizations \cite{huang}:
\begin{equation}
\langle n_{\vec{k}}\rangle = \frac{2}{e^{\beta \hbar \omega} -
1}\label{e1}
\end{equation}
Let us specialize to a scenario in which all the photons have nearly
the same energy so that we can write,
\begin{equation}
\langle n_{\vec{k}}\rangle = \langle n_{\vec{k'}} \rangle \delta (k
- k'),\label{e2}
\end{equation}
where $\langle n_{k}'\rangle$ is given by (\ref{e1}), and $k \equiv
|\vec{k}|$. The total number of photons $N$, in the volume $V$ being
considered, can be obtained in the usual way,
\begin{equation}
N = \frac{V [k]}{(2 \pi)^3} \int^\infty_0 dk4\pi k^2 \langle
n_k\rangle\label{e3}
\end{equation}
where $V$ is large. Inserting (\ref{e2}) in (\ref{e3}) we get,
\begin{equation}
N = \frac{2V}{(2 \pi)^3} 4\pi k^{'2} [\epsilon^\Theta - 1]^{-1} [k],
\Theta \equiv \beta \hbar \omega ,\label{e4}
\end{equation}
In the above, $[k] \equiv [L^{-1}]$ is a dimensionality constant,
introduced to compensate the loss of a factor $k$ in the integral
(\ref{e3}), owing to the $\delta$-function in (\ref{e2}): That is, a
volume integral in $\vec{k}$ space is reduced to a surface integral
on the sphere $[\vec{k}] = k'$, due to our constraint that all
photons have nearly the same energy.\\
We observe that, $\Theta = \hbar \omega /KT \approx 1$, since by
(\ref{e2}), the photons have nearly the same energy $\hbar \omega$.
We also introduce,
\begin{equation}
v = \frac{V}{N}, \lambda = \frac{2 \pi c}{\omega} = \frac{2 \pi}{k}
\, \mbox{and} \, z = \frac{\lambda^3}{v}\label{e5}
\end{equation}
$\lambda$ being the wavelength of the radiation. We now have from
(\ref{e4}), using (\ref{e5}),
$$(e - 1) = \frac{vk^{'2}}{\pi^2} [k] = \frac{8\pi}{k'z} [k]$$
Using (\ref{e5}) we get:
\begin{equation}
z = \frac{8 \pi}{k'(e-1)} = \frac{4 \pi}{(e-1)} [k]\label{e6}
\end{equation}
From (\ref{e6}) we conclude that, when
\begin{equation}
\lambda = \frac{e - 1}{4} = 0.4[L]\label{e7}
\end{equation}
then,
\begin{equation}
z \approx 1\label{e8}
\end{equation}
or conversely. We are working in the cgs system, so that the units
in (\ref{e7}) are $cm$. As pointed out, this could be further confirmed from other points of view.\\
We next observe that the author's 1997 cosmology used the ubiquitous zero point energy to deduce a model of an accelerating universe with a small cosmological constant (unlike in its previous 1960s version of Zeldovich and others) which lead to the famous cosmological constant problem \cite{weinberg}.\\
At this time the ruling cosmological paradigm was exactly the opposite -- a dark matter dominated decelerating universe \cite{mg8,ijmpa,ijtp}. It has also been known that the Zero Point Field leads to the Lamb shift \cite{bd,it}. It is quite remarkable that radiations due to the Lamb shift have a frequency of around 1000 megacycles, which falls right within the microwave region to a few centimeters radiation. So there is a sea of microwave photons in the universe whose frequencies differ by small amounts. By the above argument these nearly mono energetic photons would condense to a value around 4 millimeters wavelength which is exactly the cosmic microwave background.

\end{document}